\documentclass[prl,twocolumn,superscriptaddress,aps,nofootinbib]{revtex4-1}
\usepackage{graphicx}
\usepackage{natbib}
\usepackage{url}
\usepackage[normalem]{ulem}
\usepackage{color}
\usepackage{amsmath,amsmath,multirow,amssymb,color}
\usepackage[T1]{fontenc}       

\begin{document}

\title{Disorder From the Bulk Ionic Liquid in Electric Double Layer Transistors}

\author{Trevor A. Petach}
\affiliation{Department of Physics, Stanford University, Palo Alto, CA 94305, USA}
\affiliation{Stanford Institute for Materials and Energy Sciences (SIMES), SLAC National Accelerator Laboratory, Menlo Park, California 94025, USA}
\author{Konstantin V. Reich}
\affiliation{Fine Theoretical Physics Institute, University of Minnesota, Minneapolis, Minnesota 55455, United States}
\affiliation{Ioffe Institute, St Petersburg, 194021, Russia}
\author{Xiao Zhang}
\affiliation{Department of Physics, Stanford University, Palo Alto, CA 94305, USA}
\author{Kenji Watanabe}
\affiliation{National Institute for Materials Science, 1-1 Namiki, Tsukuba 305-0044, Japan}
\author{Takashi Taniguchi}
\affiliation{National Institute for Materials Science, 1-1 Namiki, Tsukuba 305-0044, Japan}
\author{Boris I. Shklovskii}
\affiliation{Fine Theoretical Physics Institute, University of Minnesota, Minneapolis, Minnesota 55455, United States}
\author{David Goldhaber-Gordon}
\affiliation{Department of Physics, Stanford University, Palo Alto, CA 94305, USA}
\affiliation{Stanford Institute for Materials and Energy Sciences (SIMES), SLAC National Accelerator Laboratory, Menlo Park, California 94025, USA}
\email{goldhaber-gordon@stanford.edu}

\begin{abstract}
  \normalsize Ionic liquid gating has a number of advantages over solid-state gating, especially for flexible or transparent devices, and for applications requiring high carrier densities. However, the large number of charged ions near the channel inevitably results in Coulomb scattering, which limits the carrier mobility in otherwise clean systems. We develop a model for this Coulomb scattering. We validate our model experimentally using ionic liquid gating of graphene across varying thicknesses of hexagonal boron nitride, demonstrating that disorder in the bulk ionic liquid often dominates the scattering.
\end{abstract}

\maketitle

\vspace{0.5cm}

Electrolyte gating has generated considerable interest as a method to induce high carrier densities in a variety of materials. \cite{Goldman2014} A voltage applied between a gate and the sample causes ions in the electrolyte to migrate to the sample surface. At the sample surface, an electric double layer is formed, which can be viewed as a capacitor with ions on one side and charge carriers on the other, with nm-scale separation. This small separation, and thus large capacitance, along with flexibility, transparency, and facile processing, enables exploration of regimes that cannot be accessed by standard solid-state metal-gate structures. Superconductivity,\cite{Ueno2008} metal-insulator transitions,\cite{Scherwitzl2010a,Nakano2012a,Jeong2013} and magnetism\cite{Shimizu2013} have been observed using electrolyte gates.

A drawback to electrolyte gating is the introduction of disorder where the electrolyte touches the channel, including electrochemical modification of the channel.\cite{Petach2014,Bubel2015,Browning2016} Separating the electrolyte from the channel using a thin layer of hexagonal boron nitride (hBN) improves cleanliness. \cite{Gallagher2015, Li2015} However, the ions in the electrolyte still create long-range Coulomb potentials, which cause unavoidable scattering. In this paper, we investigate the effect of this scattering on carrier transport in electric double layer transistors.

We focus on ionic liquids, since their wide electrochemical stability windows and low vapor pressures have led to their widespread use in electric double layer transistors. \cite{Sato2004} We use graphene channels since they have exceptionally high room-temperature mobility, \cite{Geim2005} which can be further increased by using hBN, rather than SiO$_2$, substrates. \cite{Dean2010} Such high mobility makes the scattering from the ionic liquid easier to observe. To investigate this scattering and experimentally test our theory, we measured transport in ionic liquid gated graphene with different thicknesses of hBN spacer and with two different ionic liquids.

\section{Results and discussion}
\subsection{Model for ionic liquid gating}
\label{sec:Model}
One commonly used model for scattering in graphene is a two dimensional layer of point charges with random in-plane positions located a small distance away from the graphene \cite{Hwang2007}. This model has successfully described the linear increase in conductance with carrier density in a number of experiments \cite{Chen2008,Browning2016}.

However, this model does not  describe scattering from an ionic liquid electrolyte, since the ions in the liquid are located at a variety of distances away from the graphene, and their positions are strongly correlated with one another. Thus, we develop a different model.

\begin{figure}[]
  \includegraphics[width=3.375 in]{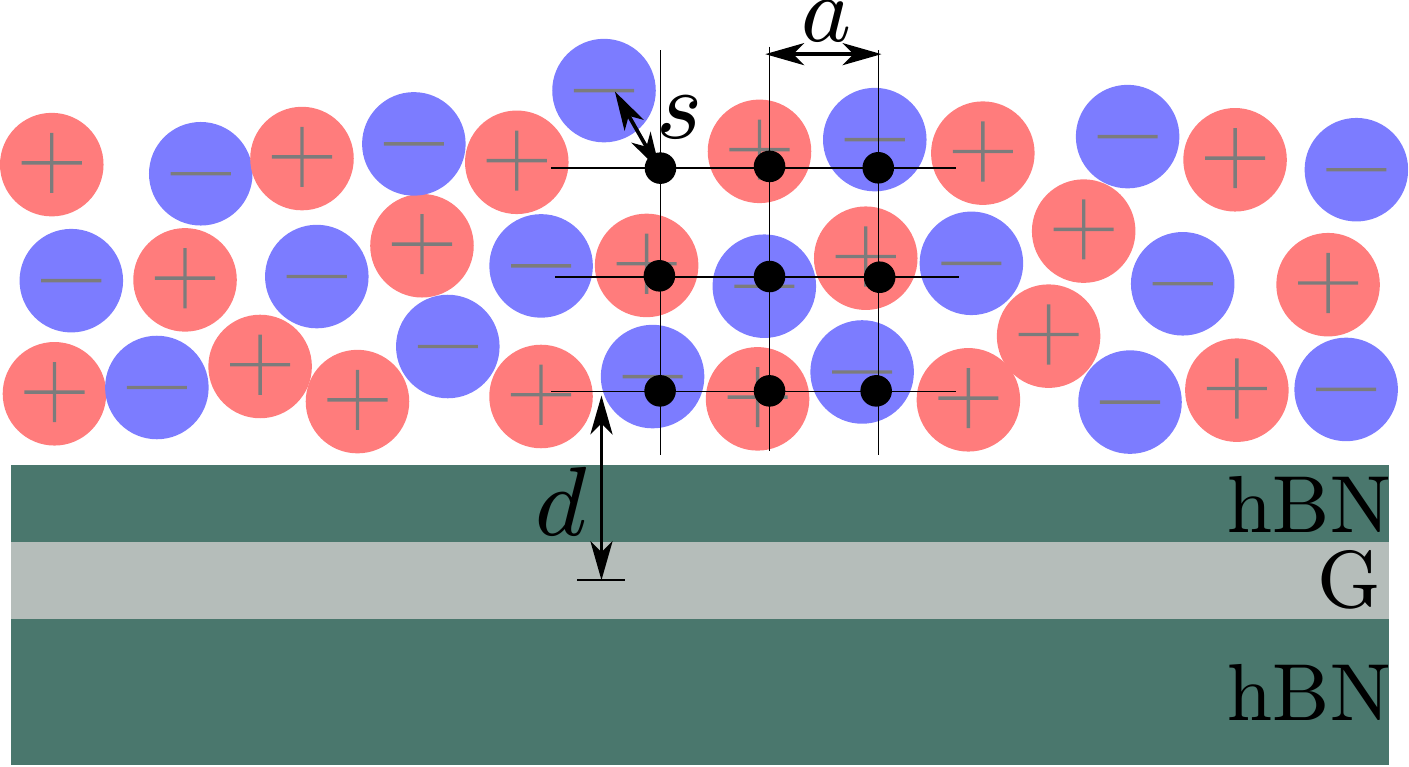}
  \caption{Schematic representation of ionic liquid gated graphene (G) covered by hexagonal boron nitride (hBN). We presume that  the centers of the  ions in the liquid resemble a periodic structure with period $a$ shown by dark dots. A small shift, $s$, of an ion from the position of a black dot produces a dipole, which scatters carriers in the graphene channel. The separation between the graphene and the first layer of ions is $d$.} 
  \label{fig:overview}
\end{figure}

At room temperature, the Coulomb interaction between neighboring ions is much larger than $k_{B}T$, so the positions of the ions are strongly correlated. They form a structure that resembles sodium chloride, with alternating anions and cations arranged periodically with spacing $a$. If the ions were arranged exactly periodically, they would not cause any scattering.\footnote{It might alter the band structure of the channel near $k \simeq 1/a$.}  Scattering of carriers in the channel is due to small shifts, $s$, of the ions from their periodic positions, as shown in Fig.~\ref{fig:overview}. These shifts can be due to both packing disorder and thermal fluctuations. Each results in a small, randomly oriented dipole with arm $s$. An ensemble of such random dipoles scatters electrons in graphene. We consider graphene sandwiched between two layers of hBN and study the dependence of the resistance on carrier concentration $n$ and the separation $d$ between the graphene and the ions. 

The potential $V$ induced by each dipole decreases with distance $r$ as $V \propto s/r^{2} $.  As we derive in Methods, the scattering from such dipoles leads to the resistivity
\begin{equation}
  R_{\rm{bulk}} = R_{0} \left(\frac{s}{a}\right)^2 \frac{1 }{k_Fa} I_b(\alpha,k_Fd),
  \label{eq:bulk}
\end{equation}
where 
\begin{equation}
  I_b(\alpha,k_Fd) = \frac{8}{3}\pi \alpha^{2} \int\limits_0^1 dy \frac{y^3 \sqrt{1-y^2}}{(y+2\alpha)^2} \int\limits_{2 y k_Fd}^{\infty} dz K_0(z)^2.
\end{equation}
Here $R_0 = h/2e^2$ is the resistance quantum, $n$ is the carrier density, and $\alpha = e^{2}/4\pi\hbar v_{F} \kappa \varepsilon_{0}$ is the Coulomb interaction parameter, where $v_{F}$ is the Fermi velocity and $k_{F}=\sqrt{\pi n}$ is the Fermi wave vector. Choosing the value for the dielectric constant is tricky because the graphene is surrounded by both hBN and ionic liquid, and hBN is an anisotropic dielectric. In the limit where the Fermi wavelength is shorter than the bottom hBN thickness and longer than the top hBN thickness, $\kappa \approx (\kappa_{\rm{IL}} + \kappa_{\rm{BN}})/2$,\cite{Landau1984} where $\kappa_{\rm{IL}}$ is the dielectric constant of the ionic liquid and $\kappa_{\rm{BN}}$ is the average of the in-plane and out-of-plane dielectric constants in hBN.

We can investigate the small separation limit. When $\alpha \ll 1$, as it is when graphene is surrounded by hBN and ionic liquid, we find that for $k_{F}d \ll 1$, $I_{b} = 2\pi^{3} \alpha^{2}/9$. Thus, the resistance does not depend on the separation $d$ for separations much smaller than the Fermi wavelength. 

To clarify the origin of the $s$ and $k_{F}$ dependence in Eq.~\eqref{eq:bulk} at $d=0$, we start from the well known expression for the contribution to the resistance from randomly distributed Coulomb scatterers. Namely, the scattering from a layer ions with concentration $n_{\rm{ion}}$ is $R \sim R_{0} n_{\rm{ion}}/n$, where $n$ is the carrier concentration.\cite{Hwang2007} One can modify this result for scattering from dipoles with 3D concentration $N \simeq a^{-3}$. First we note that the dipole potential is $s/r$ times smaller than Coulomb potential, and scattering is mostly due to those nearby dipoles within a distance approximately equal to the separation between carriers, which is $\sim n^{-1/2} \sim k_{F}^{-1}$. Using this distance as the effective $r$, the scattering rate for dipoles is $(k_{F} s)^{2}$ times smaller than for the Coulomb potential. Second, instead of $n_{\rm{ion}}$ we use two dimensional concentration $N k_F^{-1}$. Combining these two results we get $R_{\rm{bulk}} \sim R_0 (k_{F }s)^2 N k_{F}^{-1}/ n  \sim R_{0} (s/a)^{2}/(k_{F} a)$. The full calculation in Eq.~\eqref{eq:bulk} agrees with this intuitive result up to the numerical factor $I_b$.\footnote{Note that the  electron wavelength $n^{-1/2}$ varies in our experiment between 5 and 15 nm. It is larger than the scale of additional structuring and layering of ionic liquids within $\sim 2$ nm of interfaces, which was discovered near strongly charged surfaces.\cite{Israelachvili_IL_review,Perkin_2010}. This  structuring does not have a large impact on our results.} 

\subsection{Ionic liquid gated graphene}
\label{sec:results}

\begin{figure}[]
  \centering
  \includegraphics[]{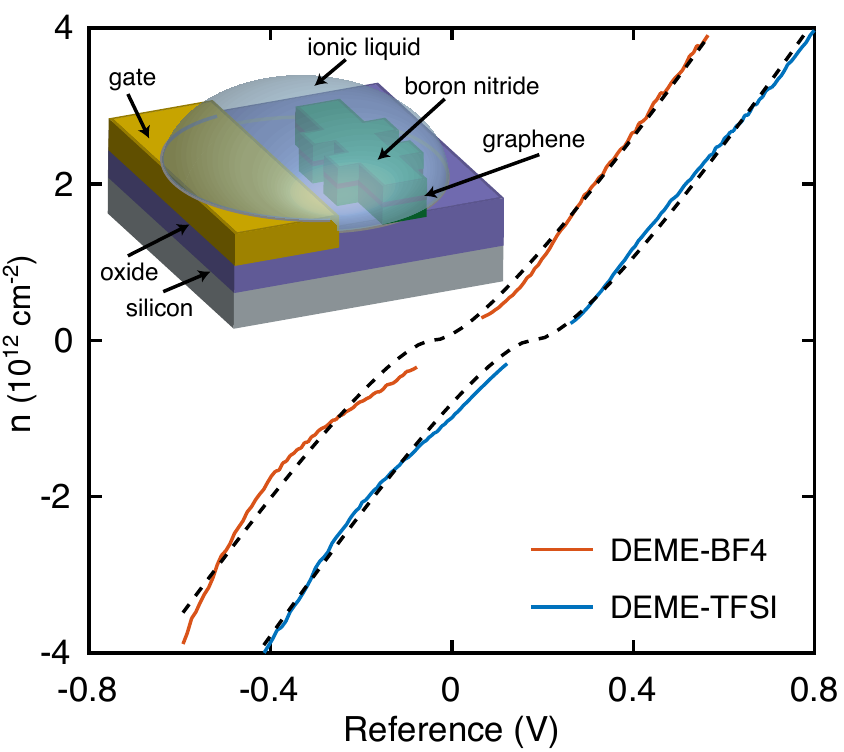}
  \caption{Carrier density in electrolyte gated graphene covered by 2 layers of hBN relative to a Fc/Fc+ reference electrode. The dashed lines are calculations presuming a constant geometric capacitance of $10^{13}$ carriers cm$^{-2}$ V$^{-1}$ in series with the quantum capacitance. \cite{Fang2007} The inset shows the device geometry, including a gate electrode and a silicon back gate.}
  \label{fig:schematic} 
\end{figure}

Our samples have hBN above and below the graphene. The bottom hBN flake is usually close to 40 nm thick. The top flake (or ``spacer''), if present, varies in thickness from 1 to 12 layers and separates the ionic liquid from the graphene. A schematic of a typical device is shown in Fig.~\ref{fig:schematic}. We used two ionic liquids, both with a diethylmethyl(2-methoxyethyl)ammonium (DEME) cation. To investigate the effect of disorder in the liquid, we used a bis(trifluoromethylsulphonyl)imide (TFSI) anion that is a similar size to DEME, forming a relatively ordered liquid, and a tetrafluoroborate (BF$_4$) anion that is smaller than DEME, forming a more disordered liquid.

The devices with hBN spacers have higher mobility than both our un-protected samples and other electrolyte-gated graphene samples in the literature. \cite{Efetov2010a,Ye2011,Browning2016} Increasing the thickness of the spacer increases the mobility, as expected.

To make a quantitative estimate of the scattering from the ionic liquid, we presume Matthiessen's rule holds, so that the sheet resistance is

\begin{equation}
R = R_{\rm{IL}} + R_{\rm{intrinsic}},
\end{equation}
where $R_{\rm{intrinsic}}$ arises from scattering unrelated to the ionic liquid, such as from phonons and crystal defects. We measured $R_{\rm{intrinsic}}$ in all samples before adding ionic liquid and found little variation from sample to sample. We presumed that this contribution to the resistance remained unchanged after adding ionic liquid.\footnote{The scattering rate from any process that depends on the dielectric constant will change when liquid is added since the dielectric constant changes. However, phonon and short-range scattering, which are the main contributions to $R_{\rm{intrinsic}}$, are mostly independent of dielectric constant.} We used Eq.~(\ref{eq:bulk}) to find the total scattering from the ionic liquid by fitting one parameter - the average relative displacement of the ions $s/a$ - to the data.

In our calculations, we used an average ion diameter of $a = 8$~\AA\ for DEME-TFSI and 7~\AA\ for DEME-BF$_4$.\cite{Sato2004} We also used $d = Nt_{\rm{BN}} + a/2$, where $N$ is the number of hBN layers and $t_{\rm{BN}} \simeq 3.4$~\AA\ is the thickness of a single layer of hBN. For the dielectric constant, we note that the in-plane and out-of-plane dielectric constants of hBN are 4 and 7. \cite{Geick1966} The dielectric constants of several liquids similar to the two we use in our study vary between 10 and 15. \cite{Wakai2005} Taking a rough average of these values, we used $\kappa = 9$.

\begin{figure}[]
  \includegraphics[width=3.375in]{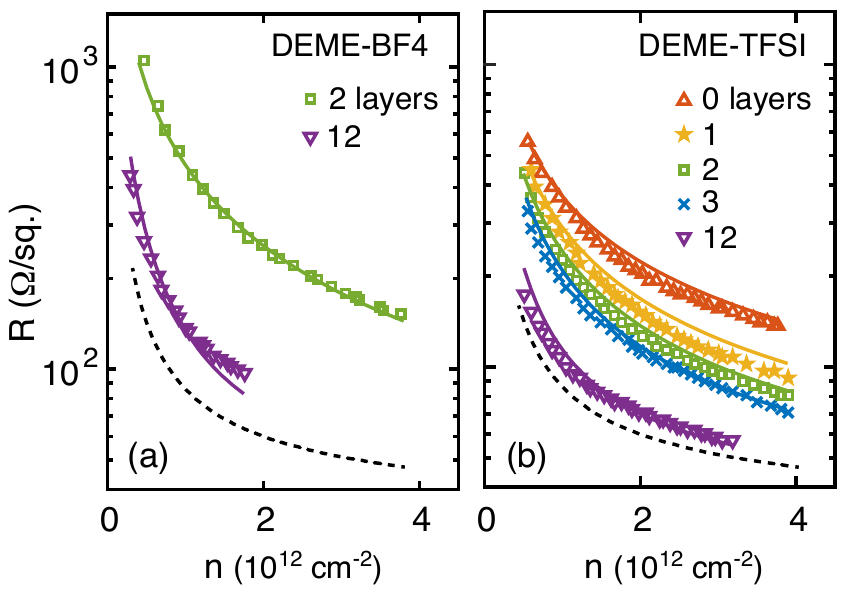}
  \caption{Sheet resistance of graphene covered by varying thicknesses of hBN using (a) DEME-BF4 and (b) DEME-TFSI ionic liquid gate. The solid lines are calculations using the fit parameter $s/a$. The dashed line is data without ionic liquid, which indicates $R_{\rm{intrinsic}}$.}
  \label{fig:thickness}
\end{figure}

In Fig.~\ref{fig:thickness}, we show that the fit agrees well with the data over a range of hBN spacer thicknesses. For DEME-TFSI and DEME-BF$_{4}$ ionic liquids we get $s/a \simeq 0.22 \pm 0.01$ and $s/a = 0.37 \pm 0.02$ respectively. Since the size difference between DEME and BF$_4$ is larger than between DEME and TFSI, we expect more disorder in the bulk liquid in DEME-BF$_4$. Indeed, the observed sheet resistances for DEME-BF$_4$ are larger than for DEME-TFSI, and the best fit value of $s/a$ is greater. We also note that $s/a > 0.1$ for both liquids, consistent with the Lindemann melting criterion, which states that in a liquid the typical shift $s$ is larger than 10\% of the atomic distance. \cite{Lindemann1910}

In the derivation of Eq. \eqref{eq:bulk} we assume that the dipoles  are  randomly  oriented  and are uncorrelated. Let us dwell here  on the  justification  of this  assumption.  The  Coulomb  interaction energy   between   neighboring   ions, $E_{C}=e^{2}/4\pi \kappa \varepsilon_{0} a$, is much larger than  room temperature, $E_{C} \simeq 10 k_{B}T$. \cite{Israelachvili_screening_IL}. On the other hand, the interaction energy between  nearest-neighbor dipoles is of the order of $(s/a)^{2}E_{C}$  and,  for  the  experimental  values $s/a \simeq 0.3$, does not exceed $k_{B}T$.  Therefore, the orientation of the  dipoles can be approximately considered as random.

\subsection{Scattering from excess surface ions}
To change the carrier density by gating, there must be an excess of one species of ion near the hBN. These ions are reminiscent of delta-doping layers in GaAs quantum wells, which are known to cause Coulomb scattering of the carriers in the two-dimensional electron gas. \cite{Pfeiffer1989,dasSarma2015} To isolate the effect of the excess surface ions, we used the silicon back gate to set the carrier density in the graphene to be different from the surface ion concentration in the liquid. 

\begin{figure}[]
  \includegraphics[width=3.375in]{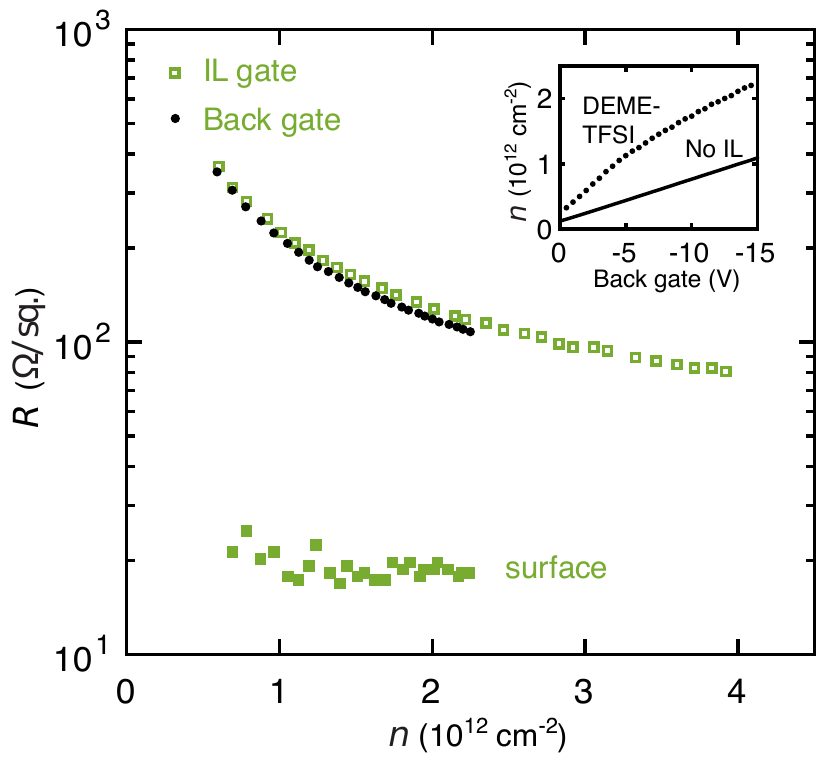}
  \caption{The sheet resistance versus carrier density using the top gate and the back gate, on a sample with 2 layers of hBN. The surface data points are the contribution to the resistance from the excess surface ions, which is the difference between the curves scaled by the fraction of carriers induced by the back gate. The inset shows the carrier density as a function of back gate voltage with and without ionic liquid. The ion density is the difference between the curves, and the carrier density is the top curve.}
  \label{fig:TGvsBG}
\end{figure}

The scattering from the bulk liquid, $R_{\rm{bulk}}$, depends only on the carrier density. It is insensitive to whether carriers were induced by the back gate or by excess ions. However, the resistance due to the scattering from the excess ions is proportional to the excess ion concentration $n_{\rm{ion}}$. It decreases when the back gate is used to induce carriers. Using the back gate to reach a certain electron density should result in less scattering than using the top gate to reach the same carrier density, since using the back gate results in lower excess ion concentration. To quantitatively estimate the scattering from excess ions, we could take the difference in resistance between carriers induced by the top gate ($R_{\mathrm{t}}$) and back gate ($R_{\mathrm{b}}$). However, there is some coupling between the back gate and the ionic liquid, so changing the back gate voltage changes both the carrier density and the excess ion concentration. By measuring the capacitance between the graphene and the back gate with no ionic liquid on the sample, we can quantify this coupling. As shown in the right inset of Fig.~\ref{fig:TGvsBG}, the excess ion concentration is equal to about half of the carrier density at a given back gate voltage, with the top gate grounded.

Presuming that the scattering rate from the excess ions is proportional to $n_{\rm{ion}}$, the contribution of the excess ions to the graphene resistance should be $(R_{\mathrm{t}} - R_{\mathrm{b}})(n - n_{\rm{ion}})/n$, which we plot using open symbols in the Fig.~\ref{fig:TGvsBG}. We see that the resistance due to excess ions at $2\times10^{12}$ cm$^{-2}$ is $\simeq 20 ~\Omega$, an order of magnitude smaller than the expected value ($\simeq 300 ~\Omega$) if we assume that excess ions are randomly positioned near the hBN surface (at $d =8$ \AA). \cite{Hwang2007} 

There are several possible reasons why the scattering from excess ions could be suppressed below the expected value. First, the in-plane position of the ions may be correlated. In this case, the scattering from the excess ions can be suppressed, depending on the structure factor of the correlations, \cite{Li2011} analogous to correlation of ionized dopants in GaAs. \cite{Buks1994} Second, the excess ions may not be located at the surface of the hBN. If the excess ions are located away from the surface, then the scattering would be reduced. The relatively small capacitance suggests that the ions may indeed be separated from the hBN surface. As shown in Fig.~\ref{fig:schematic}, the geometric capacitance is about $10^{13}$ cm$^{-2}$ V$^{-1}$ for both liquids. Presuming that the excess ions form a simple parallel plate capacitor with the graphene, the ratio $d/\epsilon = 5$ \AA. Using $\epsilon = 9$ as before, $d = 45$ \AA, which would reduce the expected contribution to scattering to approximately the observed contribution. Regardless of the mechanism, it seems that the excess surface ions contribute little to scattering.

\section{Conclusions}
\label{sec:conclusions}

In agreement with other work, we have shown that separating the ionic liquid from the channel in an electric double layer transistor using a thin hBN spacer improves the mobility. However, we have also shown that there is always some scattering arising from the ions in the ionic liquid. A simple model that captures the disorder in the bulk liquid due to small displacements of the ions from a periodic structure can explain the observed transport properties.

A more comprehensive experiment and analysis may yield detailed information about the location of the ions in the double layer and the bulk, their correlation functions, and their behavior as a function of potential.

Already, we have seen that there is a trade-off between achieving high carrier densities, which requires thin hBN spacers, and achieving high mobilities, which requires thick spacers. With no spacer, we have shown that there is an upper limit to the mobility that can be achieved with ionic liquid gates.

\section{Methods}
\label{sec:methods}

\subsection{Derivation of Model Results}
\label{sec:derivation}
So long as the mean free path, $v_{F} \tau$, is much larger than the Fermi wavelength, the resistance per square of graphene can be rewritten in terms of the scattering time, $\tau$, as\cite{Ando2006}
\begin{equation}
  R=R_{0} \frac{1}{k_F v_{F} \tau}.
  \label{eq:resistance_def}
\end{equation}
The scattering time is \cite{DasSarma2011}

\begin{equation}
  \frac{1}{\tau} = \frac{2\pi}{\hbar} \int \left(\frac{d^2k}{(2\pi)^2} N(z)dz \left| \frac{V(q,z)}{\epsilon(q)} \right|^2   F(q) (1-\cos \theta)  \delta(\varepsilon(k_{F}) - \varepsilon(k)) \right). 
  \label{eq:tau}
\end{equation}
Here $z$ is the perpendicular distance from the ion to the graphene, $q = 2k_F \sin(\theta/2)$ is the scattering vector, $N(z)$ is the volume concentration of scattering centers in the ionic liquid, $V(q, z)$ is the Fourier transform of the scattering potential, $\epsilon(q)$ is the dielectric function, $(1 - \cos \theta)$ is the transport factor for graphene that prohibits back scattering, $F(q)$ is the form factor for the carrier wavefunctions, and $\varepsilon(k)$ is the energy dispersion relation. In graphene, $\epsilon(q) = (1 + 4\alpha k_F /q)$, $\varepsilon(k)=\hbar v_{F} k$, and $F(q)=(1+\cos(\theta))/2$.

So we find:

\begin{equation}
  \label{eq:RR}
  R=16 \pi R_{0} \alpha^{2} \int \limits_{0}^{1} \frac{y^{4} \sqrt{1-y^{2}}}{(y+2\alpha)^{2}} \int \limits_{d}^{\infty} N(z) dz \left( \frac{V(2 k_{F} y ,z) 4\pi \kappa \varepsilon_{0}}{2\pi e^{2}} \right)^{2}
\end{equation}

In this derivation, we have assumed that the electron gas is degenerate, which is a valid assumption for densities above $10^{11}$ cm$^{-2}$ in graphene at room temperature. All measurements in our experiments were made above this density. Thus, for a given $s$ the  scattering  rate  does not  depend  on temperature.  At  the  same  time,  it is likely that $s$ increases  with   temperature and, therefore, resistivity  increases with $T$ as well.

In the case of bulk disorder, each shifted ion contributes to the scattering, so $N(z) \simeq a^{-3}$. The dipole potential is $V(\rho,z)=e^{2} s \cos (\phi) /4\pi \varepsilon_{0} \kappa (\rho^{2}+z^{2})$, where $s$ is the displacement of the ion, and $\phi$ is the angle between the axis of the dipole and the position of an electron. The average distance between electrons, $n^{-1/2}$, is greater than the thickness of the top hBN. From an electrostatic point of view, we can neglect the top hBN and assume that the graphene is placed between the ionic liquid and the bottom hBN. The effective dielectric constant in this case is
$(\kappa_{\rm{IL}} + \kappa_{\rm{BN}})/2$.

After averaging over all possible dipole orientations, we find

\begin{equation}
  \label{eq:dipole_Fourier}
  |V(q,z)|^{2}= \frac{1}{3} \left(\frac{2 \pi e^2 s}{4\pi \varepsilon_{0} \kappa} K_0(qz)\right)^{2},
\end{equation}
where $K_0$ is the modified Bessel function of the second kind.

Evaluating the integral in Eq.~\eqref{eq:RR}, we arrive at Eq.~(\ref{eq:bulk}).

\subsection{Fabrication of graphene heterostructures}

We fabricated hBN-graphene-hBN stacks using a dry transfer method with PPC/PDMS stamps and deposited them on highly doped silicon substrates with 300 nm of oxide. \cite{Wang2013a} We used e-beam lithography (30 kV, 250 - 300 $\mu$C/cm$^2$, 1:3 isopropanol:water developer) with 950k PMMA resist, followed by reactive ion etching in a mixture of 10\% O$_2$ and 90\% CHF$_3$ to define the hall bars (Oxford PlasmaLab 80, 50 W, 150 mTorr). We e-beam evaporated 5 nm Cr and 75 nm Au to make contacts and annealed the final devices in 10\% H$_2$ in Ar at 350 C for 1 hr. We found that sweeping an AFM tip across the surface in contact mode (50 nN) after annealing further increased the mobility, and we did this treatment on all samples. After sweeping, the mobilities were close to the phonon limit. \cite{Hwang2008a} We fabricated a Pt gate electrode nearby and dropped a small amount of ionic liquid (baked overnight at 80 C in high vacuum) over the gate and the hall bar. 

\subsection{Transport measurements}

 All measurements were performed at 300 K in high vacuum ($< 10^{-5}$ Torr). We used a Stanford Research Systems SR830 lockin amplifier to measure 4-terminal resistance and Hall voltage. The current sourced was 1 $\mu$A, and the Hall bars were 2 $\mu$m by 3 $\mu$m. We calculated carrier density from the difference in the Hall voltage at $+$20 mT and $-$20 mT, which was applied using a solenoid wound around the vacuum chamber. We used a Keithley K2400 Sourcemeter to apply gate voltages. The leakage currents were less than 1 $\mu$A/cm$^2$.

 We measured the applied potential relative to a Fc/Fc+ reference electrode. We dissolved a small amount of ferrocene (Fc) and ferrocenium (Fc+) directly in the ionic liquid ($<1$ mg/mL) and measured the potential on an immersed Pt wire using a high impedance ($>$ 10G$\Omega$) voltmeter. At a given gate voltage, the potential at the Fc/Fc+ reference is about 300 mV less than the potential at a Ag/Ag+ reference in 1 M AgNO$_3$, which was formed by dissolving AgNO$_3$ in ionic liquid in a separate tube that was separated by 5 $\mu$m diameter fritted glass from the rest of the cell.

 \section{Acknowledgement}
 We thank Shu Hu and Nathan Lewis for providing valuable insight and lab assistance to develop the reference electrode used in this work. The experimental work was supported by the Department of Energy, Laboratory Directed Research and Development funding, under contract DE-AC02-76SF00515. The theoretical work was supported by the National Science Foundation through the University of Minnesota MRSEC under Award Number DMR-1420013. Part of this work was performed at the Stanford Nano Shared Facilities, supported by the National Science Foundation under award ECCS-1542152. T. P. was supported by the Department of Defense through a National Defense Science and Engineering Graduate Fellowship and by a William R. and Sara Hart Kimball Stanford Graduate Fellowship. K.V.R was supported by the Russian Science Foundation under grant 17-72-10072. K.W. and T.T. acknowledge support from the Elemental Strategy Initiative conducted by the MEXT, Japan and JSPS KAKENHI Grant Numbers JP15K21722.

 \providecommand{\latin}[1]{#1}
 \makeatletter
 \providecommand{\doi}
  {\begingroup\let\do\@makeother\dospecials
  \catcode`\{=1 \catcode`\}=2 \doi@aux}
\providecommand{\doi@aux}[1]{\endgroup\texttt{#1}}
\makeatother
\providecommand*\mcitethebibliography{\thebibliography}
\csname @ifundefined\endcsname{endmcitethebibliography}
  {\let\endmcitethebibliography\endthebibliography}{}

\end{document}